# Long term experience in Autonomous Stations and production quality control

**L. Lopes**[b,*], **A. B. Alves**[f], **P. Assis**[a], **A. Blanco**[b], **N. Carolino**[b], **M. A. Cerda**[c,e], **R. Conceição**[d], **O. Cunha**[b], **C. Dobrigkeit**[g], **M. Ferreira**[a], **P. Fonte**[b,d], **L. de Almeida**[f], **R. Luz**[a], **V. B. Martins**[f], **L. Mendes**[a], **J. C. Nogueira**[a], **A. Pereira**[b], **M. Pimenta**[a], **R. Sarmento**[a], **V. de Souza**[f], **B. Tomé**[a]

[a] *Laboratório de Instrumentação e Física Experimental de Partículas (Lip),*
*Av. Professor Gama Pinto, n. 2, Complexo Interdisciplinar (3is), 1649-003 Lisboa, Portugal*

[b] *Laboratório de Instrumentação e Física Experimental de Partículas (Lip),*
*Departamento de Física da Universidade de Coimbra, 3004-516 Coimbra, Portugal*

[c] *Observatório Pierre Auger,*
*Malargüe, Argentina*

[d] *Coimbra Polytechnic - ISEC, Coimbra, Portugal*

[e] *Instituto Nazionale di Fisica Nucleare sezione di Roma Tor Vergata,*
*Roma, Italy*

[f] *Instituto de Física de São Carlos, Universidade de São Paulo, Av. Trabalhador São-Carlense 400,*
*São Carlos, Brasil*

[g] *Instituto Nazionale di Fisica Nucleare sezione di Roma Tor Vergata,*
*Roma, Italy*

*E-mail*: `luisalberto@coimbra.lip.pt`

ABSTRACT: Large area arrays composed by dispersed stations are of major importance in experiments where Extended Air Shower (EAS) sampling is necessary. In those dispersed stations is mandatory to have detectors that requires very low maintenance and shows good resilience to environmental conditions. In 2012 our group started to work in RPCs that could become acceptable candidates to operate within these conditions. Since that time, more than 30 complete detectors were produced, tested and installed in different places, both indoor and outdoor. The data and analysis to be presented is manly related to the tests made in the Auger site, where two RPCs are under test in real conditions for more than two years. The results confirm the capability to operate such kind of RPCs for long time periods under harsh conditions at a stable efficiency. In the last years Lip and USP - São Carlos start collaboration that aim to install an Eng. Array at BATATA (Auger) site to better study and improve the resilience and performance of the RPCs in outdoor experiments. The organization of such collaboration and the work done so far will be presented.

---

[*] Corresponding author.

## Contents

## 1. Introduction

Resistive Plate Chambers [2] can be found in a large number of experiments, mainly in indoor but also in outdoor environments [3-6]. Six years ago [1, 8] we started research on the development of a chamber to operate outdoors with residual maintenance, low power and gas consumption and we believe to be now close to a robust and mature detector solution.

In the framework of the Pierre Auger Observatory upgrade, RPCs have been proposed as a dedicated muon detector to better estimate the muonic component of Extensive Air Showers (EAS), further constraining the nature of the cosmic rays and hadronic interactions taking place in the EAS development. An engineering array will be installed in the infill region. Although this array will only allow the collection of a limited sample of moderate energy cosmic ray showers, it will be of extreme importance to set a calibration point as it allows a direct measurement of the muonic component of the showers. Furthermore, the Engineering array will be installed in the same region as AMIGA, an underground muon detector, allowing to cross-calibrate the two detectors. The instrumentation of 7 tanks as described in [8], will allow to study and improve the performance of Resistive Plate Chambers (RPCs) in outdoor inhospitable environment.

In this work we present the outdoor results taken over more than one year in the infill region. These results sustain the stable performance of the RPCs detectors in outdoor conditions. In the second part we will describe the organization related to the R&D, production and test of the engineering array detectors in collaboration with the colleagues from the São Carlos Physics Institute, Brazil.

## 2. Long term Outdoor results

The environmental conditions at the Pierre Auger Observatory situated on the vast plain known as the *Pampa Amarilla* (yellow prairie) in the western Argentina [7], are well know and described at [9]. A detailed description of the detector module can be found within the same reference. The detectors operates with a mono-component gas R-134a, at a flow rate of 4 cc/min.

To assure a stable efficiency over the time was implemented the automatic HV adjustment, to compensate any temperature and/or pressure variations with high voltage. A detailed reading of reference [9] will clarify the previous sentence. There, is shown the importance of the automatic HV adjustment to keep the reduced electric field "constant", which is of major importance to keep the gain constant and consequently the charge spectrum and efficiency at stable values.

The setup used to monitor the efficiency in real field (Tierra del Fuego, figure 1) conditions consists of two detector modules placed inside a precast structure underneath one water Cherenkov tank. This way we can use the tank and one RPC to due the trigger and study the other RPC performance. Besides supporting the tank, the precast structure protects the RPC modules from direct nature environment and help on filtering the electromagnetic component of the shower.

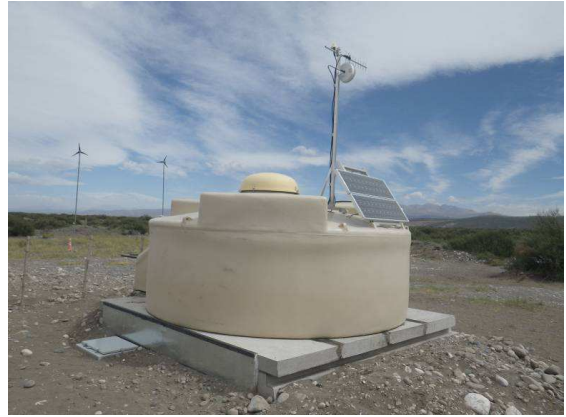

**Figure 1:** Picture of the Tierra del Fuego tank. The two RPC modules (not visible) are inside the precast structure underneath the tank.

The frontend electronics doesn't give charge information, just count the number of times each pad triggers within a certain time window. This way is not possible to use the charge to be sure about the gain stability, the reduced electric field is the only controllable variable that will allow to operate at a stable gain and consequently (hopefully) at a stable efficiency. For these propose, a permanent monitoring of temperature and absolute pressure is mandatory to adjust the HV (each 15 minutes) and keep E/N as constant as possible.

The two RPCs were turned on in November 2015 and in March 2016 the automatic HV adjustment was started. Figures 2 and 3 clear shown the capability to operate at a stable E/N only through the automatic HV adjustment to correct from temperature and/or pressure variations. To a better understanding of the behavior of the variables shown in figures 2 and 3 is better to look also for the equation that describes the dependence of the reduced electric field with temperature, absolute pressure and effective voltage.

$$\frac{E}{N} = 0.0138068748 \times \frac{V_{eff,Volts}}{d_{cm}} \frac{(T_{°C} + 273.15)}{P_{mbar}}, \quad [Td]$$

In first approximation, due to the low currents drawn by the chambers is possible to neglect the contribution of the potential drop in the glass electrodes and consider that all applied voltage will be available within the gap, this is $V_{eff} \cong HV$.

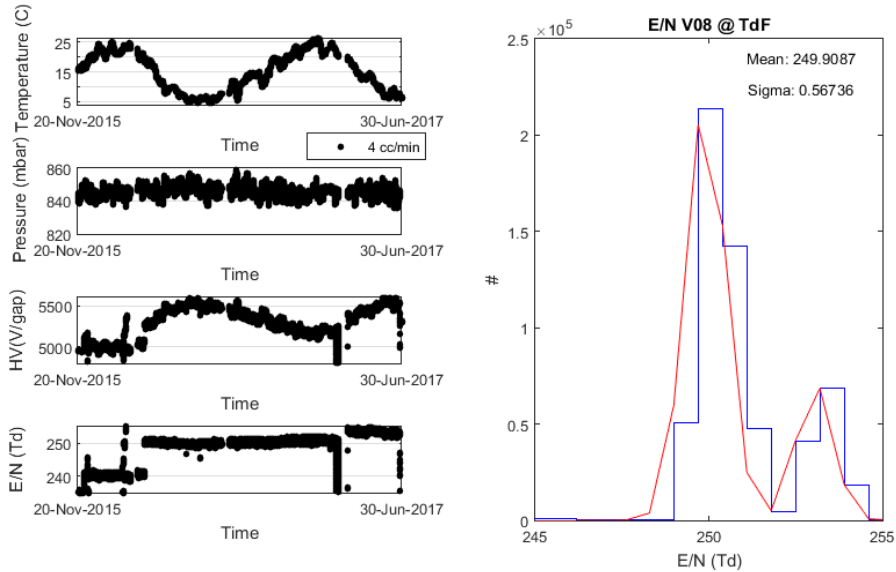

**Figure 2:** RPC on the trigger. Most part off the time at 250 Td as reference E/N, the rest at 254 Td. On the left side the reduced electric field and the three variables considered for its determination via the automatic adjustment of the applied high voltage. This detector operates at a gas flow rate of 4 cc/min. On the right side the distribution of the reduced electric field since March 2016.

In both figures is clear the seasonal variation of temperature and the inverse effect that causes in the HV. Despite the large temperature daily excursions in the site the large inertia from tank, precast and soil strongly smooth the effect of temperature in the detectors, reducing by a factor 10 [9] the daily temperature gradient. In these temperature conditions the HV adjustments will be very small, just a few tens of Volts at most at each step. This is also important because in the adjustment process the effect of temperature in the electrodes is not taken into account. For the situations where is necessary to correct from large (more than 10 ºC) temperature gradients is impossible to keep assuming a negligible effect of the glass (or any resistive electrode) resistivity in the potential drop and consequently in the effective voltage. This is even more significant if the operation temperature is above the room temperature. In that situations the background and leakage currents increase faster than the glass resistivity decreases causing an increase in the drop potential, that reduces the effective potential resulting in a lost of gain. For temperatures between zero and room the situation should be better once the background rate is real low and the leakage currents are absent. Even the increase by one order of magnitude in the glass resistivity will not produce a considerable effect on the potential drop, once is compensated by the very low currents drawn by the chamber.

With respect to absolute pressure the daily and seasonal excursions are below 5% at the most, that corresponds at a maximum of 50 V adjust in the HV. This small variations in the HV for a constant temperature will produce negligible background and leakage currents, making the absolute pressure much less important than temperature for situations where is applied the automatic HV adjustment. Is clear for both detectors that is possible to operate at a stable E/N and as result expected a stable ("constant") efficiency, which is of main importance in experiments where Extended Air Shower sampling is necessary.

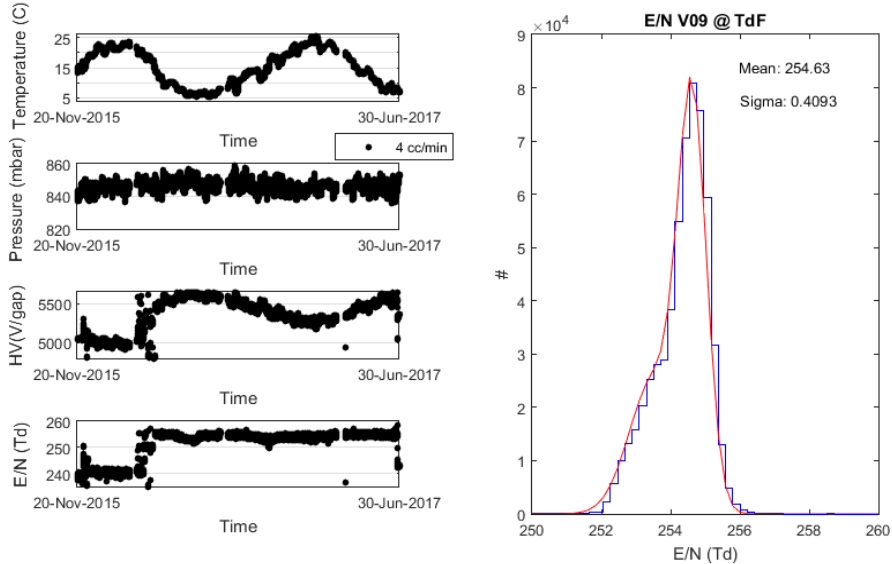

**Figure 3:** RPC under study. The chosen reference value for E/N was 255 Td. On the left side the reduced electric field and the three variables considered for its determination via the automatic adjustment of the applied high voltage. This detector operates at a gas flow rate of 4 cc/min. On the right side the distribution of the reduced electric field since March 2016.

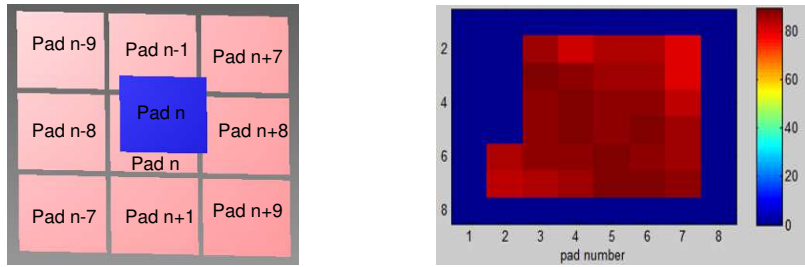

**Figure 4:** Left side. Scheme of the efficiency measure setup. The pad n of the chamber under study can be crossed by the muons that cross de same pad number and all neighbour ones of the chamber in the trigger. Right side. Efficiency over the area, all measure pads shows very similar efficiency values around 85%. Which is an acceptable value for chamber with two gaps of 1 mm at an absolute pressure off 850 mbar and room temperature.

The two RPCs are separated by 36 mm and each one is cover with 64 signal pickup pads. The trigger is defined by a coincidence between one chamber and the water Cherenkov tank. One efficient event is when we have a hit in the pad *n* off the RPC under study and in the same pad *n* of the RPC on the trigger or in any neighbor pad as in figure 4, left side. Due to efficiency definition all the border pads can't be take into account. The four missing pads in the right side of figure 4 are due to dead channels. In reference [9] is proven that both RPCs had similar efficiency dependence with E/N. There is also explained why is not possible to reach the same efficiency plateau as in the laboratory. This is due to different operation pressures, at TdF site the pressure is 15% lower that at sea level as in the laboratory. Correct this 15% drop in the pressure only by the increase in the high voltage will not be enough. Once the pressure reduction also affect the detector gain through the gas density. To compensate from this effect is necessary to increase the gas thickness, increasing the gap(s) width or the number of gaps. In other words is necessary to keep constant the ratio between the number of ionization clusters and the number of ionization steps.

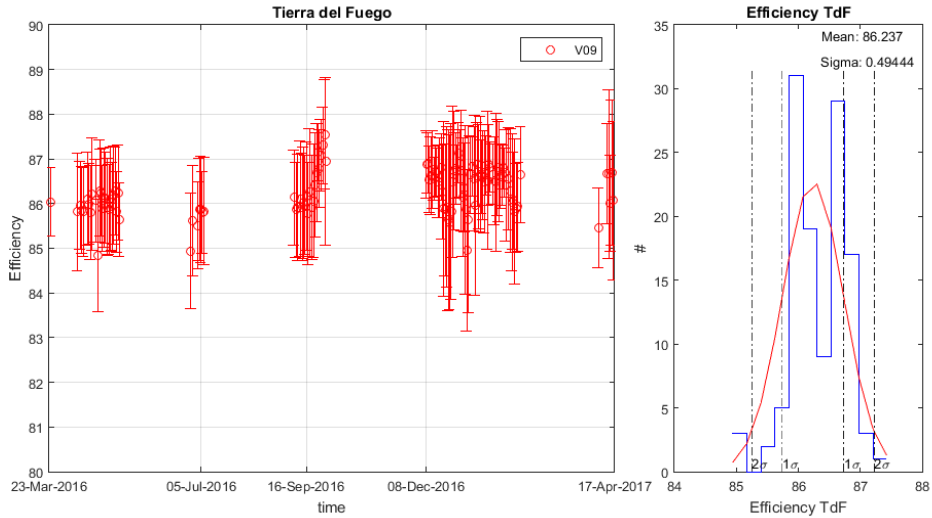

**Figure 5:** On the left side, the efficiency stability over time (point/day) for more than 1 year at a "constant" E/N of 254 Td, see figure 3. On the right side, the distribution of the efficiency, that proves the stability over the time. The two peaks could be addressed to different HV power supplies offsets, once around September 2016 had to be changed due to malfunction.

Figure 5 summarizes the field results over more than a year. Is clear that is possible to operate trigger RPC chambers in standalone remote stations with low gas consumption at a stable efficiency. The electronics used in this setup were not prepared to work in outdoor harsh conditions and during the data taking period experiment some problems and unfortunately in the middle of 2017 suffer a major malfunction and we had to stop data taking. New electronics, prepared for this specific harsh operation conditions is already done and tested and will be installed in the field within this year.

## 3. Engineering Array

Although the Auger Collaboration did not opt for the use of RPCs to the upgrade, it was considerer important to continue the study and development of RPCs for astrophysics. In this sense in a Portugal-Brazil collaboration was decide to instrument 7 Auger tanks with the design proposal for the Auger upgrade. The tasks were distributed between both institutes. The project, sensitive volume and support systems were developed and delivered by Portugal. The structural, shielding and "tight" aluminum box, the pickup pads and RPC module integration were developed and assemble in Brazil. The sensitive volume is very well detailed in reference [9]. Before sending the sensitive volumes to Brazil they need to get positive evaluation in two tests:

Argon discharge, were is possible to clean and check the gaps uniformity; efficiency and charge maps at the working point with R-134a. In figure 6 is the correlation between the current and HV with Argon in the gaps. At NPT (see level and room lab room temperature) conditions, close to 3000 V the discharge becomes permanent over all area and is possible to explore the linear dependency to extract the glass resistivity. Below 3000 V the proportional regime dominates as expected. The observation of this dependencies and the determination of the glass resistivity are the first step to confirm the correct chamber assembly. The chamber is then leave at 3000 V for at least 3 days to "clean" the gap surfaces. This way when change to Tetrafluorethane the conditioning period will be shorter.

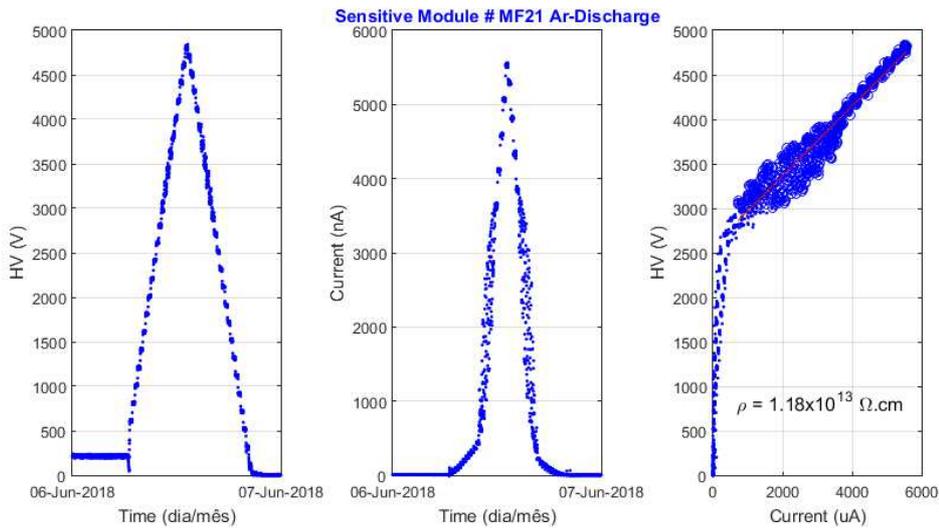

**Figure 6:** The Argon discharge quality test. With test is possible to check the gap uniformity, clean the gap surfaces and had a sustained evaluation of the detector assembly quality. Start the application of HV with a inert gas also prevent from some possible dust polymerization and reduces the conditioning period when change to tetrafluorethane.

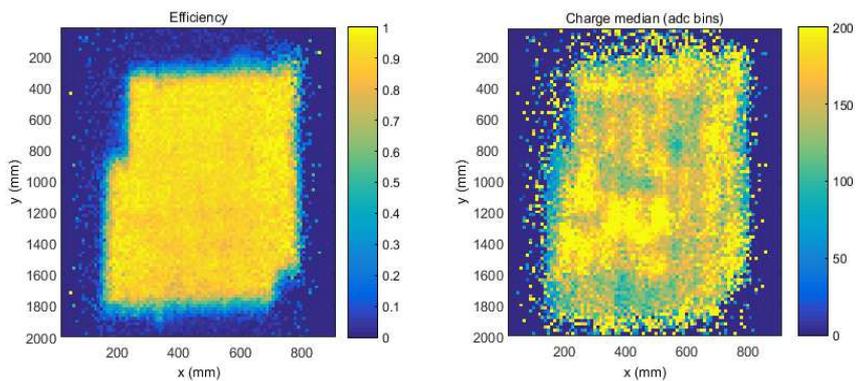

**Figure 7:** Efficiency and charge median 2D plots. These two plots are the final quality control test to approved each sensitive volume before being sent to Brazil. At $E/N \cong 238$ Td, the efficiency should be above 90% over all the area. The charge median should be above 100 adc units. Although, the charge measurement is to much sensitive to the signal induction, which turns in less uniformity over the area.

After the Argon test the sensitive module is placed in a muon telescope and start to be flushed with Tetrafluorethane. The muon telescope allows to test 3 detectors simultaneously. For each sensitive module we monitor the background current and rate, temperature, pressure and relative humidity. After the replacement of the Argon by the Tetrafluorethane, the HV is turned on and with the current limit to 1 uA we start the HV rump up until reach the working point (middle of the efficiency plateau). At NPT and at a gas flow rate of 10 cc/min these process normally takes between 2-3 days. Then the automatic HV adjustment is turned on to keep the reduced electric field "constant" around 238 Td, and start data taking. In figure 7 the two important plots to considerer the sensitive volume approved: the 2D plots of the efficiency and charge median. As can be seen the efficiency is close to 100% over all the scanned area (5 pads are missing due limited electronic channels). The charge median in adc bins is also quite uniform over all area. Although due to detector module construction and signal induction, the charge measurement is more sensitive to small variations as can be seen in the corresponding plot.

The first 20 sensitive volumes are already in Brazil been integrated inside the aluminum boxes. Production and test of the remain 20 is in progress and should be ended within July 2018. The supporting subsystem as gas distribution and monitoring systems, HV power supplies and frontend electronics are also under production and some units are already in Brazil to be integrated in the detector modules. All these subsystem are very low power consumption, and were entirely developed and assembly by Lip.

In the Physics Institute of São Carlos our colleagues start the integration of the first 10 detector modules in the second half of 2017. In figure 8.a is the 3d cad of the structural/shielding aluminum box. The sensitive volume, pad plan and environmental sensors are assemble inside. This volume is then sealed in order to minimize contamination with moisture and dust. The volume receives the gas output from the sensitive volume to increase the removal of any moisture. A second/small aluminum box is assemble on the top of the first one to housing all the electronic subsystems. Also this box should be sealed (as much as possible), see the picture in figure 8.b. The bubbler block has 3 monitoring columns to follow the gas flow rate, temperature and relative humidity in all the 3 volumes described above. In the end each detector module will be a closed aluminum structure with 5 connections, 3 form communications, 1 for power and one for gas feed. The monitoring and safety columns of the bubbler block can be seen/check by naked eye to better assist local staff during any gas checking or bottle exchange.

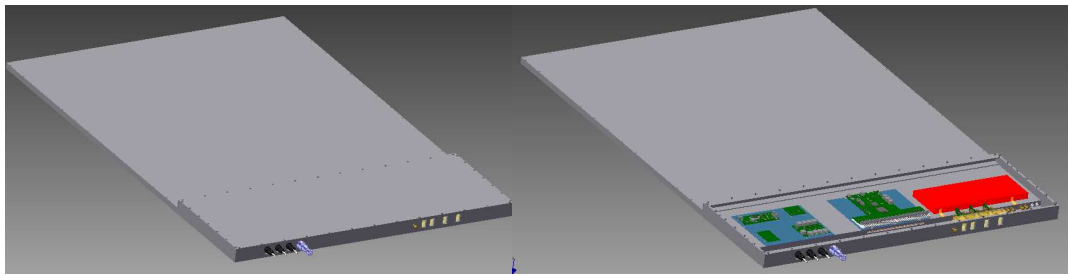

**Figure 8a:** 3D cad of the aluminium shielding box were the sensitive module and pad plan are assembled. The small extra volume will housing the subsystems as: HV power supply (red); gas monitoring bubbler block (yellow), Maroc frontend board (centre green board); LV power supply, multiplexer and gas system communication (left green boards).

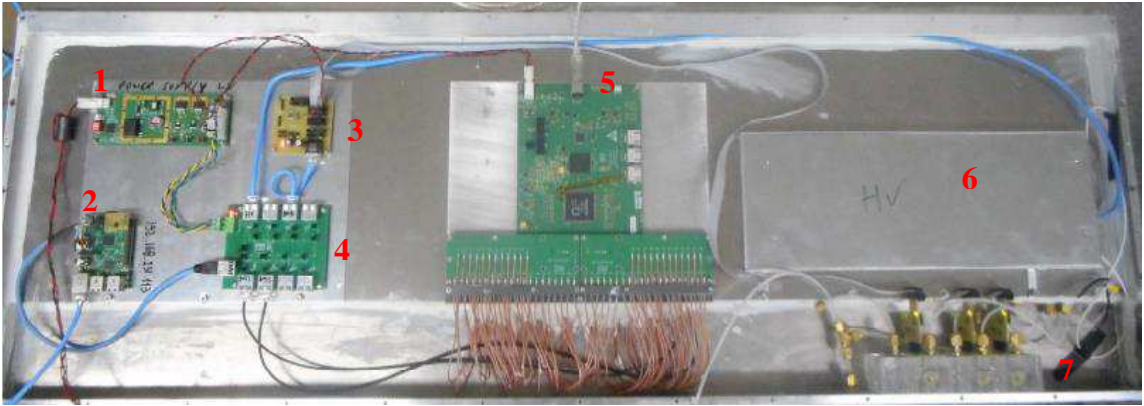

**Figure 8b:** Picture of the subsystems already installed in a detector module. 1. LV power supply, 2. raspberry pi, only used for test, 3. gas system communication board, 4. I2C multiplexer, 5. Maroc board, 6. HV power supply and 7. gas monitoring bubbler block. Together all these subsystems requests less than 10 Watts.

The first actions in the field to prepare the precast structures to support the tanks and housing the RPC modules already start at Malargue. The first field installations of the Engineering Array are scheduled for the last quarter of this year.

## 4. Conclusion

Since 2012, when start the development of RPCs for outdoor applications we experience some difficulties. Most of them only discovered when start tests in the field. During this process we also learn that our experience for indoor/controlled environmental is important but really far from been enough for this challenge. Despite all these barriers is clear that is possible to use RPCs for outdoor, standalone applications. Under harsh operation conditions, with very low gas flow rate at a stable efficiency. This is important for experiments where EAS sampling is necessary, because the need to cover large remote areas. To improve the robustness and resilience of the detectors is very important to monitoring all ambient variables and prepare the system in the way to minimize as much as possible there influence in the detector operation.

The possibility to install an Engeniring Array with seven station in the infill region of the Auger site was embraced by us together with our Brazilian colleagues as a unique opportunity to improve the performance of RPCs for outdoor and also their possible contribution for the physics understanding in EAS experiments. The collaboration between Portugal and Brazil in the development, construction, installation and operation of this array of stations should be consider as an important step in the RPCs technology grown in South America. Which is a place where many outdoor experiments are based and many others are under study. The first 10 detector modules should be complete assembled in Brazil in the next months and the first installations in the field at Malargue are schedule for the last quarter of this year.

## Acknowledgments

This work is supported by Portuguese national funds OE, FCT-Portugal, CERN/FIS-PAR/0023/2017 and OE, FCT-Portugal, FAPESP/19946/2014. The author R. Luz thanks the FCT-Portugal and IDPASC Portugal for the grant PD/BD/113488/2015.